\documentstyle[12pt,epsf,epsfig]{article}

\author{Steven Abel$^a$ and Toby Falk$^{b}$\\
{\small $^a$Theory Division, Cern 1211, Geneva 23, Switzerland}\\
{\small $^b$Department of Physics, University of Wisconsin,}\\
{\small Madison, WI 53706, USA} }

\title{Charge and colour breaking in the Constrained MSSM} 
\newcommand{\hepph}[1]{{\tt hep-ph/#1 }}
\newcommand{\phrd}[3]{{{\it Phys.~Rev.}~{\bf D#1} (#3) #2}}
\newcommand{\plb}[3]{{{\it Phys.~Lett.}~{\bf B#1} (#3) #2}}
\newcommand{\prd}[3]{{{\it Phys.~Rev.}~{\bf D#1} (#3) #2}}

\newcommand{\npb}[3]{{{\it Nucl.~Phys.}~{\bf B#1} (#3) #2}}

\newcommand{\leqsim}{\,\raisebox{-0.6ex}{$\buildrel < \over \sim$}\,}
\newcommand{\geqsim}{\,\raisebox{-0.6ex}{$\buildrel > \over \sim$}\,}
\newcommand{\be}{\begin{equation}}
\newcommand{\ee}{\end{equation}}
\newcommand{\ba}{\begin{eqnarray}}
\newcommand{\ea}{\end{eqnarray}}

\newcommand{\ie}{\mbox{\em i.e.~}}
\newcommand{\eg}{\mbox{\em e.g.~}}

\newcommand{\nn}{\nonumber}

\def\gev{\,{\rm GeV }}

\def\mcha{m_{\chi^\pm}}

\def\mchi{m_{\tilde\chi}}
\def\m12{m_{1/2}}
\def\tb{\tan\beta}
\def\ohsq{\Omega_\chi h^2}

\begin{document} 

\maketitle

\begin{abstract} 
\noindent We show that the physical minimum of the Constrained MSSM is
only free from dangerous charge and colour breaking minima
in the region of parameter space bounded
by $110 \leqsim \m12 \leqsim 400\gev$ and $80 \leqsim
m_0 \leqsim 170\gev$. In the remaining regions the cosmology 
is severely constrained.
\end{abstract} 
\vspace{0.5 cm}
\vfill
\begin{flushleft}
{\tt CERN-TH/98-322\\
  MADPH-98-1083 \\
  hep-ph/9810297}
\end{flushleft}
\vspace{3.0in}
\newpage

\section{Introduction}

Unphysical charge and colour breaking (CCB)
vacua~\cite{ccb1,dilaton,baer,riotto,quasi,us0,us,recent} have come under
renewed scrutiny recently\footnote{In Refs.\cite{quasi,us} the bounds
which will be of most interest here were, in deference to historical
precedent, referred to as Unbounded From Below (UFB) bounds. This is a
confusing misnomer since the directions are not, in most cases,
unbounded from below, and we revert to calling them CCB bounds.}. For
a number of models it has been found that CCB vacua are present in the
whole of the parameter space which has not already been excluded by
experiment. This is true for models where supersymmetry breaking is
driven by the dilaton~\cite{dilaton}, for $M$-theory in which
supersymmetry breaking is driven by bulk moduli fields~\cite{us,recent} and
for the MSSM at the low $\tan\beta$ fixed point~\cite{quasi}.

In this letter we extend the analysis of Ref.\cite{quasi} to a
complete determination of CCB bounds in the Constrained MSSM (CMSSM).  
The present work is partly an update of the
results of Ref.\cite{baer}, and from that study and also
Ref.\cite{quasi} it is clear that the parameter space is bounded from
different directions by CCB bounds, dark matter bounds and
experimental bounds.  Dark matter bounds (the requirement that
neutralino dark matter be consistent with a universe that is 12 Gyr
old) tend to eliminate regions where $m_0$ is large. On the other
hand, experiment (in particular Higgs and chargino searches) is
eliminating regions of low $\m12$. CCB bounds however tend to
`favour' regions where $\m12\leqsim m_{0}$ or in other words high
$m_0$ and low $\m12$. What we shall demonstrate in this letter is
that there is only a small amount of remaining parameter space in the
CMSSM which does {\em not} have global CCB minima.

Before continuing, we emphasise that there are various schools of
thought regarding CCB minima. This is because the tunneling rate from
the physical vacuum into any global CCB minima is extremely small, so
that the physical vacuum is essentially stable on the lifetime of the
observable universe.  The authors of Ref.\cite{baer}, for example,
invoke a principle that `the cosmological constant is zero in the {\em
global} minimum' in order to explain the vanishing cosmological
constant, thus requiring that the standard model minimum be deeper
than any CCB minima.  If this were the case, then CCB minima would
indeed be a severe problem and would impose severe constraints on the
MSSM.  By contrast, some other authors take a minimalist approach; if
there is no chance of tunneling, then CCB minima are not a problem.
However, one has to explain how the physical vacuum was chosen over
the much wider CCB minima~\cite{riotto,us0}. In fact this does naturally
occur in some models, but not in others (\eg supergravity models
which possess a Heisenberg symmetry \cite{BG}, including no-scale
models of supergravity \cite{gmo}).  As the vacuum choice depends on
unknown details of our cosmological history (\eg the inflationary
potential) we think that CCB minima should ultimately be regarded as a
constraint on early cosmology rather than particle physics.  We also
emphasize that since tunneling between vacua is so slow as to be
irrelevant, it is more appropriate to flag models which contain any 
CCB minima, regardless of whether they are global or local.

We begin by describing the most dangerous directions, assuming the 
usual $R$-parity invariant superpotential of the MSSM,
\be
W_{MSSM}=h_U Q H_2 U^c + h_D Q H_1 D^c + h_E L H_1 E^c+\mu H_1 H_2,
\ee
and a degenerate pattern of supersymmetry breaking with universal
scalar mass$^2$ parameters ($m_0^2$), trilinear couplings ($A_0$) and
gaugino masses ($\m12$) at the GUT or Planck scale.  The notation
is the same as in Ref.\cite{quasi}.

As emphasised in Ref.\cite{us0}, the flat directions in the MSSM are a
direct result of the adoption of $R$-parity, to prevent the proton
decaying. The dangerous $F$ and $D$ flat directions which will be of
interest in this letter are constructed from gauge invariants
involving $H_2$~\cite{carlos,tony}.  This is because its mass squared
parameter, $m_2^2$, appears in the potential along these directions,
and it must be negative in order to drive electroweak symmetry
breaking.  The first example of this kind in the literature is (see
Komatsu in Ref.\cite{ccb1})
\be
L_i Q_3 D_3 \mbox{ ; } H_2 L_i
\ee 
where the suffices on matter superfields are generation indices. 
With the following choice of VEVs;
\ba
\label{komkom}
h_2^0             &=& -a^2 \mu/h_{D_{33}} \nonumber \\
\tilde{d}_{L_3}=\tilde{d}_{R_3} &=& a \mu/h_{D_{33}} \nonumber \\
\tilde{\nu}_i   &=& a  \sqrt{1+a^2} \mu/h_{D_{33}}, 
\ea 
the potential along this direction is $F$ and $D$-flat, and
depends only on the soft supersymmetry breaking terms;
\be
\label{softv}
V=\frac{\mu ^2}{h_{D_{33}}^2} a^2 (a^2 (m_2^2+m_{L_{ii}}^2) + 
m_{L_{ii}}^2+m_{d_{33}}^2+m_{Q_{33}}^2 ).
\ee 
At large values of $a\gg 1$ the potential is governed by the first
term. Because $m_2^2$ is required to turn negative during the
renormalisation group running (for successful electroweak symmetry
breaking) the potential can develop a charge and colour breaking
minimum at a scale of $few\times \mu /h_{D_j} $. Ensuring that this
does not happen leads to the constraint in which we are interested.

The above is not quite, but is very close to, the deepest `fully
optimised' direction~\cite{ccb1,baer}. To get the fully optimized
condition we parameterize the VEV of $\tilde{\nu}_i$ as~\cite{ccb1}
\be
\tilde{\nu}_i   = \gamma_L a^2 \mu/h_{D_{33}}. 
\ee
Minimisation of $V$ with respect to $\gamma_L$ then gives 
\be 
\gamma_L^2 = \frac{1+a^2}{a^2} -
\frac{2 m_{L_i}^2 h_b^2 }
{\hat{g}^2 a^4\mu^2 }
\ee
where $\hat{g}^2 = ({g'}^2+g_2^2)/2 $, and a potential of 
\be
\label{softv2}
V=\frac{\mu ^2}{h_{D_{33}}^2} a^2 (a^2 (m_2^2+m_{L_{ii}}^2) + 
m_{L_{ii}}^2+m_{d_{33}}^2+m_{Q_{33}}^2 ) - \frac{m_{L_i}^4}{\hat{g}^2}  .
\ee 
When $\gamma_L^2<0$ then one should set $\gamma_L^2=0$ to get the most
stringent condition although, for the regions of parameter space of
interest here, this will never be the case.  The potential typically
has a depth of $\geqsim 10^{6} m_W^4 $ at the minimum.  In addition,
as shown in Ref.\cite{us0}, the depth of the minimum is actually
extremely sensitive to the choice of soft supersymmetry breaking
parameters, so that, for most reasonable choices of parameters, the
difference between Eq.(\ref{softv}) and Eq.(\ref{softv2}) is
negligible. In Ref.\cite{us0}, emphasis was also placed on the
closeness of the usual condition (that the physical vacuum be the
global minimum) to the more relevant condition (that the physical
vacuum be the only minimum).  There is generally an extremely thin,
but cosmologically interesting, region of parameter space between the
`allowed' and `disallowed' regions, where there is a CCB minimum which
is {\em local}.

In order to obtain the bound we now need to take account of the
renormalisation group running of the mass-squared parameters between
the weak and GUT scales. To do this we shall assume that the largest
mass, and therefore the appropriate scale at which to evaluate the
parameters is $\phi=h_{U_{33}} \langle h_2^0 \rangle $. This minimises
the top quark contributions to the effective potential at one-loop.
Further corrections to the potential are assumed to be small.  As
shown Ref.\cite{baer}, this approximation is adequate for determining
CCB bounds on the supersymmetry breaking parameters (although they
also note a curious bifurcation in behaviour when one varies the scale
at which the parameters are evaluated).

In the above potentials, $ \langle h_2^0 \rangle= -a^2 \mu /
h_{D_{33},E_{33}} $ so that the Eq.(\ref{softv}) is of the form 
\be
V=\frac{M_{GUT}^2}{h_{U_{33}}^2} \hat\phi
 \left( \hat\phi A + B/b \right)
\ee 
where $A=m_2^2(\phi)+m_{L_{ii}}^2(\phi)$, $B$ is the other combination 
of mass-squared parameters (also evaluated at $\phi$) which appears 
in the potentials above,
\be 
\hat\phi=\phi/M_{GUT}
\ee
and 
\be
\label{b}
b(\phi) = \frac{M_{GUT}h_{D_{33}}}{h_{U_{33}} \mu}
\ee
for the $LQD,~LH_2$ direction described above, or
\be
\label{b2}
b(\phi) = \frac{M_{GUT}h_{E_{33}}}{h_{U_{33}} \mu}
\ee
for the equally dangerous $LLE,~LH_2$ direction.  The traditional
(no global CCB minima) bound is saturated by $V=V'=0 $; the non-trivial 
solution is therefore also a solution to $\tilde{V}=\tilde{V}'=0$ where
\be
\tilde{V}=
\hat\phi A + B/b.
\ee 
Hence, only the GUT scale parameters plus the parameter $b$ enter the
traditional bound as was pointed out in Ref.\cite{us0}. In fact the
bound always becomes {\em more} restrictive with increase in $b$,
since this decreases the positive contribution to the potential.

\section{The analysis}

\subsection{CCB Minima}

First we shall examine the properties of the CCB minima when we vary
the soft supersymmetry breaking parameters.  We will not however
calculate the tunneling rates which is generally a complicated
numerical task.  In this instance, it is made particularly difficult
by the presence of more than one gauge invariant involved in the flat
directions.  (As an aside we note that, for the directions involving
only the single $UDD$ operator examined in Ref.\cite{riotto}, a very
accurate analytic approximation can be found for the tunneling rate.)
However, from dimensional considerations and estimates of tunneling
rates in Refs.\cite{riotto,us0}, it is clear that the lifetime for
these directions will be much longer than the age of the universe.

We can make additional observations on the stability of physical
vacuum by adopting the approach of Ref.\cite{baer} in the light of
Ref.\cite{us0}.  In the latter it was pointed out that the CCB bounds
are least restrictive when $A_0=-\m12$ \footnote{In our sign
conventions, the 3-4 element of the neutralino mass matrix is $-\mu$,
and the mixing term in the stop mass matrix is
$m_t(A_t+\mu\cot\beta)$.}. Hence in Fig.~\ref{fig:vevs} we show
contours of the VEV of the CCB minimum in the $m_0$, $\m12$
parameter space, for $\tb=2,10$ and $\mu<0$, and taking $A_0=-\m12$
and $0$.  Here we have plotted $\log_{10}({\rm VEV}/246\gev)$ for the
$LLE$,~$LH_2$ direction.  Above the solid contour there are no global CCB
minima. In the thin shaded region (approximately 10\gev wide) above the solid contour, 
there are  CCB minima which are only local, and 
the physical minimum is still the global minimum. 
In Ref.\cite{us} a general survey was made of all the
possible dangerous directions in field space, and this direction was
found to give the severest bounds (at least for the CMSSM) so
henceforth we shall only be considering this. The figure illustrates
that the VEVs of CCB minima are smallest when the CCB bound is close
to being saturated. This leads to the rather counter-intuitive fact
that the physical vacuum is least stable in these regions. When the
bound is strongly violated the VEVs become very large and the lifetime
of the physical vacuum increases.  The CCB bounds are least
restrictive when $A\approx - \m12$, and they are close to the analytic
approximation found in Ref.\cite{us0}; if we define $\tilde{m}_0=
m_0/\m12$ then the analytic approximation is given by
\be 
\tilde{m}_0^2 > \frac{f(3 \tilde{m}^2_0) - g (3 \tilde{m}^2_0)
(1-\rho_p)}{4-3\rho_p}
\ee
where 
\ba
f(x) &=& 1.43 - 0.16 x + 0.02 x^2 \nn\\
g(x) &=& 2.94 - 0.2 x + 0.02 x^2 \nn\\
1/\rho_p &=& 1+3.17 (\sin^2 \beta - \sin^2\beta^{QFP} ).
\ea
These functions were evaluated in Ref.\cite{us} for $\m12 = 200\gev $
and in the one loop approximation. However we use the full
two loop value of $\tan\beta^{QFP}$ which is $\approx 1.6$.  (Note
that the important factor here is the distance from the fixed point
which is given by $\rho_p$. For a given value of $\tan\beta $ this
obviously depends sensitively on the fixed point value, $\tan\beta
^{QFP}$, so we cannot expect the analytic approximation to be better
than $\sim 15$\%.) For $\tan\beta = 2,10$ and $A_0=-\m12$ we find the
bounds $m_0 \geqsim 130,40 $ respectively. The first compares
favourably with the bounds at $m_{1/2}=200\gev $ found numerically in
Fig.~\ref{fig:bnds} but the $\tan\beta=10 $ bound is found to be
larger than the analytic estimate. This suggests that the bottom quark
Yukawa and/or the two loop contributions to the beta functions are
already contributing significantly to the bounds at $\tan\beta=10$.

\subsection{Experimental Constraints}

\begin{figure}
\vspace*{-2.3in}
\begin{minipage}{6.0cm}
\hspace*{-1in}
\epsfig{file=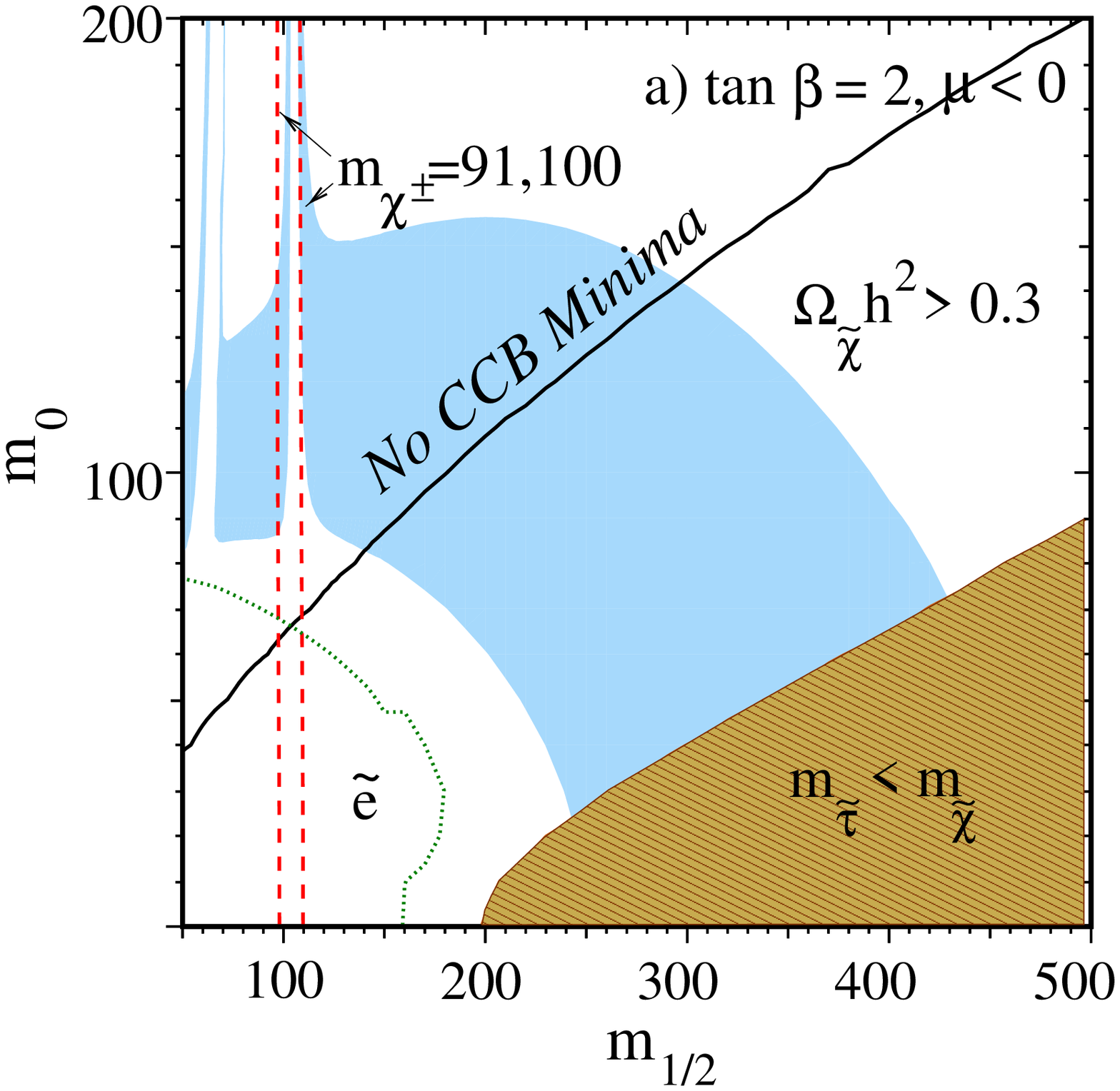,height=6in} 
\end{minipage}
\hspace*{0.3in}
\begin{minipage}{6.0cm}
\epsfig{file=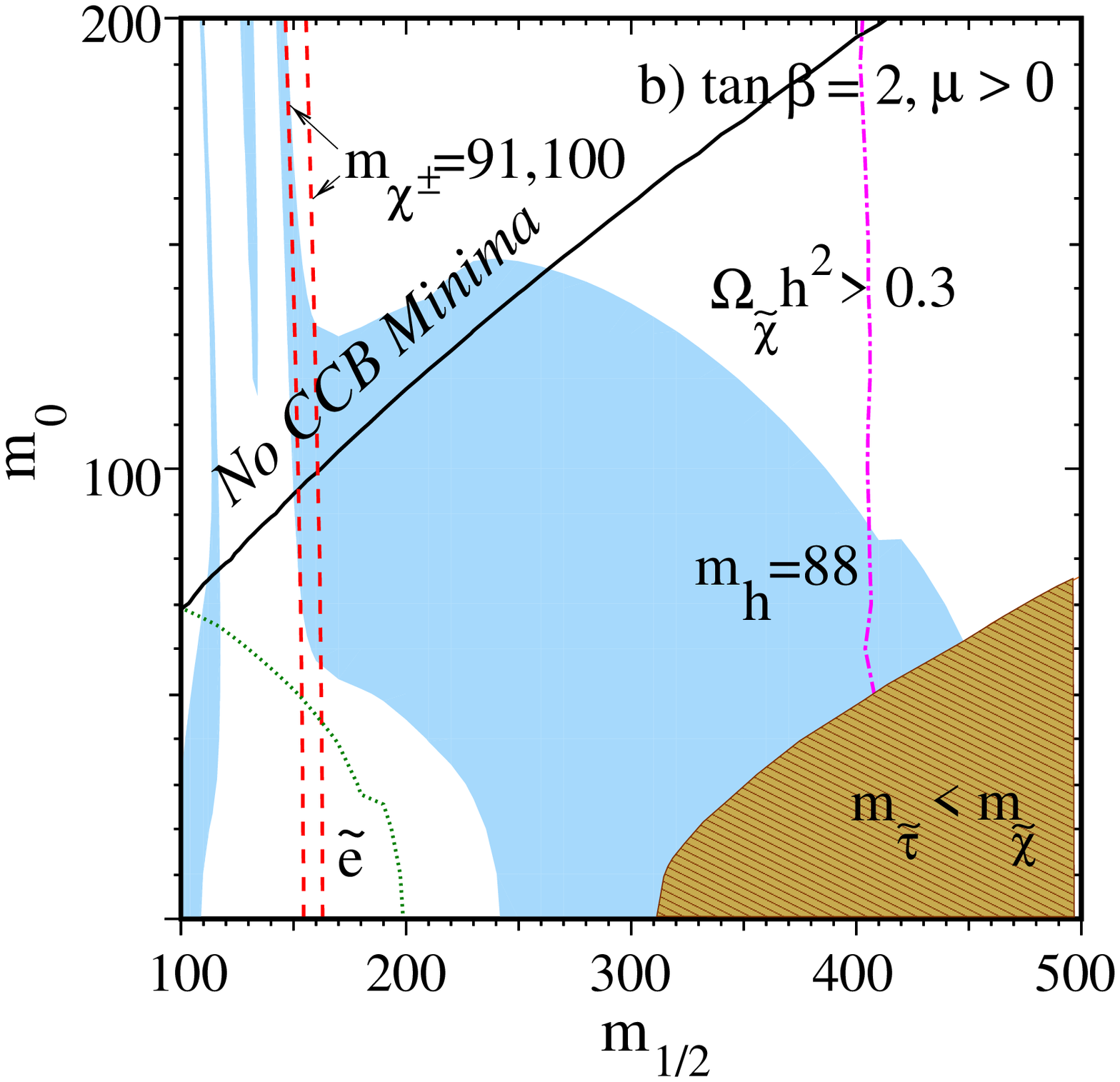,height=6in} 
\end{minipage}\hfill
\vspace{-2.0in}
\begin{minipage}{6.0cm}
\hspace*{-1in}
\epsfig{file=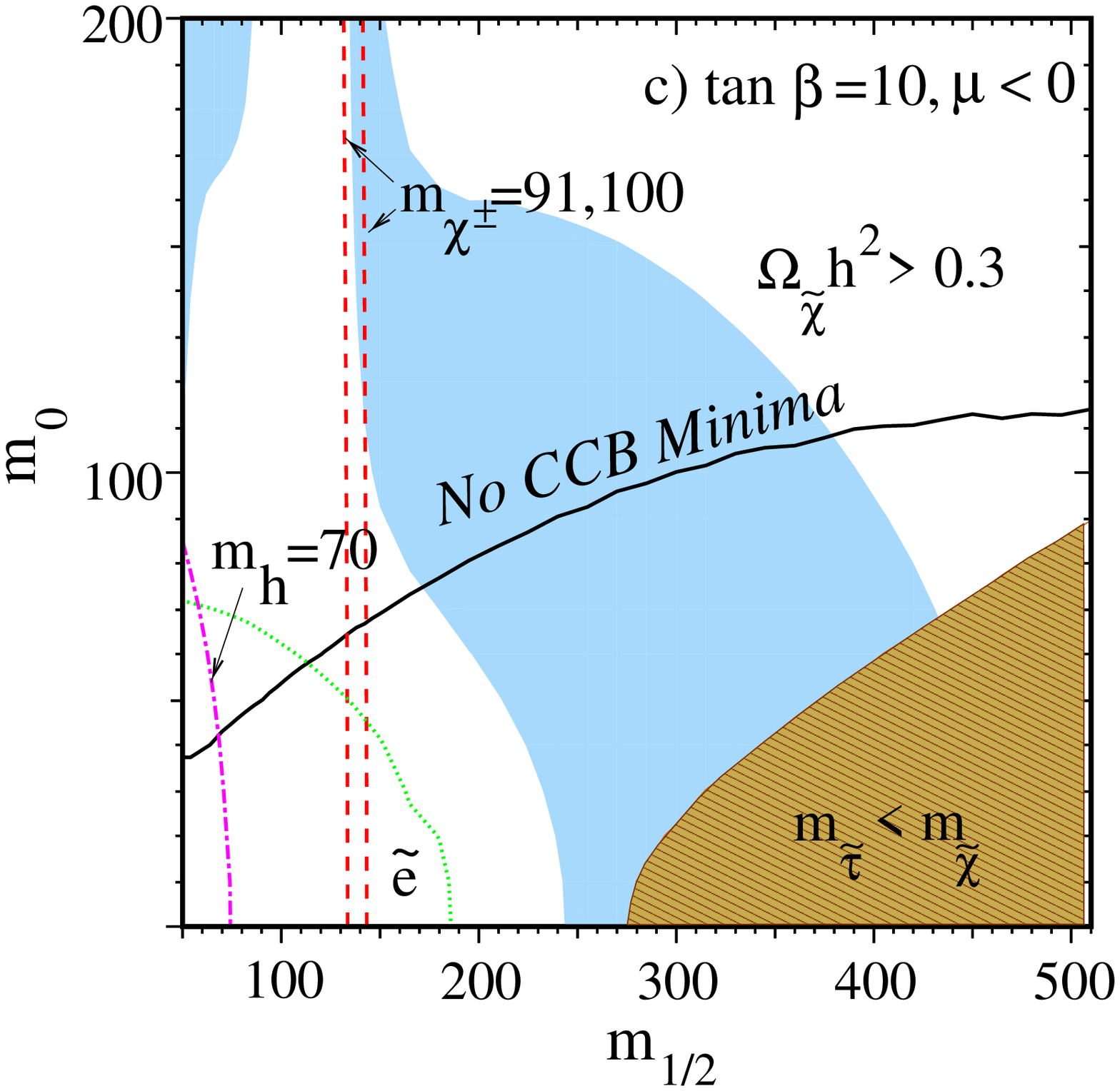,height=6in} 
\end{minipage}
\hspace*{0.3in}
\begin{minipage}{6.0cm}
\epsfig{file=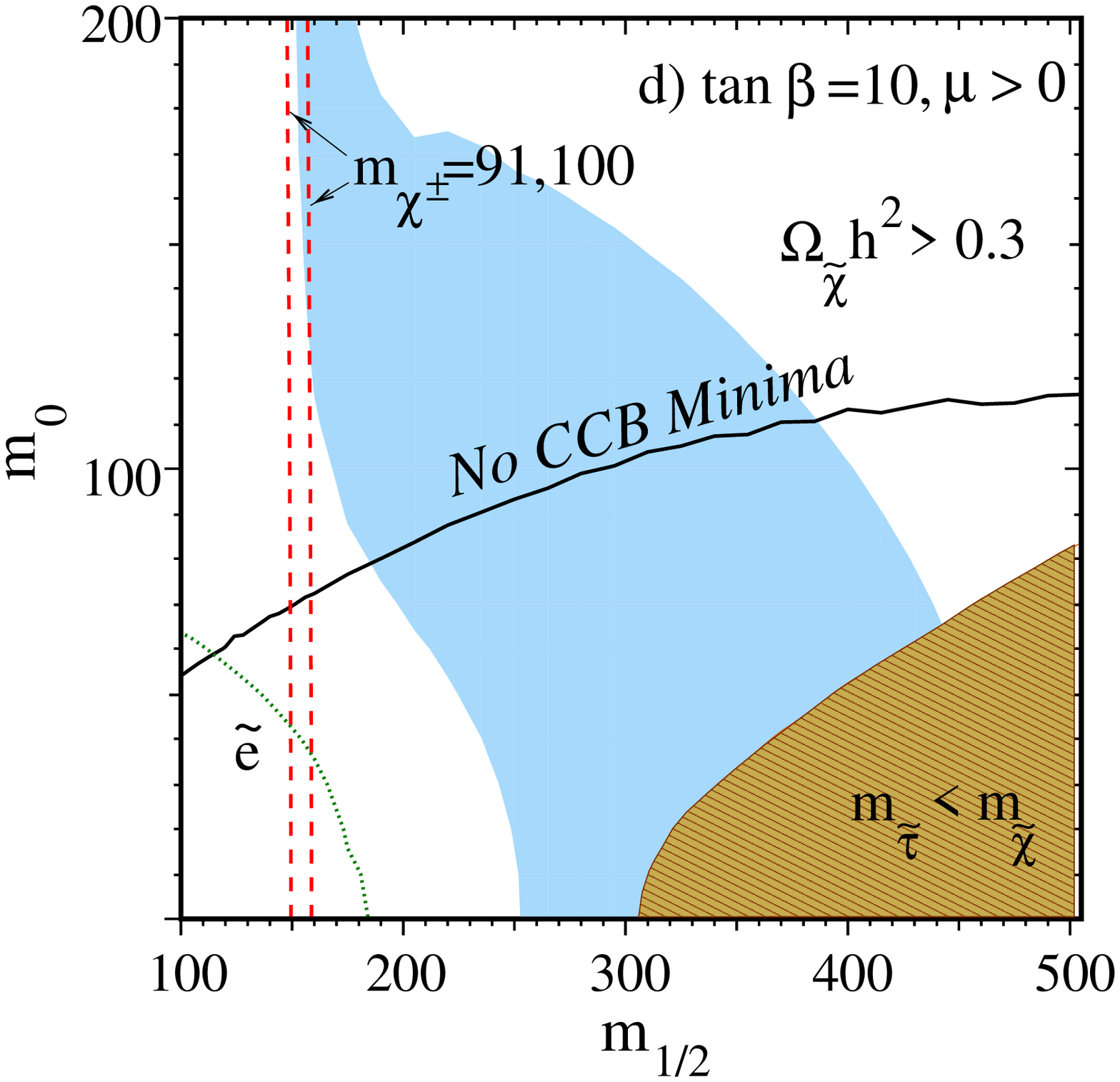,height=6in} 
\end{minipage}\hfill
\vspace{-0.7in}
\caption{\label{fig:bnds}The combined cosmological and experimental
constraints on the constrained MSSM, for $\tb=2,10$ and both $\mu<0$
and $\mu>0$.  The dashed contours represent current and future LEP
chargino bounds, dotted contours are slepton bounds, and dot-dashed
contours are Higgs bounds.  The light-shaded region gives
$0.1<\ohsq<0.3$.  Below the solid contour, CCB minima are present in
the $LLE, LH_2$ direction.  We have chosen $A_0=-\m12$ to minimise the
area containing CCB minima.}
\end{figure}

We now discuss the experimental and cosmological bounds on the CMSSM
parameter space.  Recent runs at LEP at center-of-mass energies of 172
and 183~\gev have excluded large areas of the CMSSM parameters space,
and subsequent runs at $\sim 190$ and 200~\gev will push the bounds
even further.  In the CMSSM, the dominant constraints at moderate to
high $\tb$ come from searches for chargino pair production, and modulo
a small loophole which can occur when the mass of the sneutrino is
close to the chargino mass, the experimental bounds saturate the
kinematic limit of $\mcha\sim91$~\gev.  Chargino iso-mass contours of
91 and 100~\gev, representing the current and projected LEP 200
chargino mass bounds, respectively, are displayed in
Fig.~\ref{fig:bnds} as dashed lines in the $\m12-m_0$ plane, for two
representative values of $\tb$ and both signs of $\mu$.  Also shown as
a dotted line is the current LEP183 slepton mass bound
\cite{alephslps}, which is roughly 84 GeV for large $m_{\tilde
l_R}-\mchi$.

At low $\tb$, the dominant CMSSM constraint comes from searches for
Higgs production at LEP183.  Not only are the experimental bounds
strongest at low $\tb$, roughly 90~\gev at $\tb=2$, but here the tree
level Higgs mass is also smallest ($m_h^{\rm tree}\approx m_Z
\cos2\beta$ for $m_A\gg m_Z$).  Radiative corrections to the Higgs
mass \cite{MSSMHiggs}, which depend logarithmically on the sfermion
masses, must then be very large, leading to strong lower bounds on the
masses of the sfermions, and in particular the stops.  However, the
extraction of the radiatively corrected Higgs mass in the MSSM has an
uncertainty of $\sim2\gev$, so we conservatively take $m_h>88$~\gev as
our experimental lower limit at low $\tb$.  The LEP183 Higgs bounds
are shown in Fig.~\ref{fig:bnds} as a dot-dashed line.  The entire
displayed region of Fig.~\ref{fig:bnds}a is excluded by the Higgs mass
constraint, while none of the displayed region in Fig.~\ref{fig:bnds}d
is in conflict with the current Higgs bound.  Of course the Higgs mass
corrections are very sensitive to $\tb$ for $\tb$ near the quasi-fixed
point, and, additionally, the experimental limit falls for $\tb>2$,
and so the Higgs constraint moves quickly to the left for $\tb>2$.
The chargino bound, on the other hand, moves to the right, and for
$100\gev<m_0<200\gev$, the two bounds together exclude $\m12\leqsim110
\gev$ for all $\tb$ and both signs of $\mu$.

\subsection{Cosmological Constraints}

Over most of the CMSSM parameter space, the lightest supersymmetric
particle (LSP) is a bino-like lightest neutralino $\tilde\chi$.
$R$-parity ensures that the LSP is stable over cosmological time scales,
and in the CMSSM, the LSP typically has a cosmologically interesting
relic density \cite{gold,susydm1}.  In much of the parameter space, in
fact, the relic abundance of neutralinos is so large that it is in
conflict with the observed age of the universe, $t_U>12$~Gyr, and the
corresponding upper limit of $\ohsq < 0.3$ can be used to exclude
large areas of $m_0$ and $\m12$, as follows.  In the early universe,
gaugino-like neutralinos annihilate predominantly via sfermion
exchange.  Increasing either $m_0$ or $\m12$ drives up the sfermion
masses, lowers the annihilation rate, and increases the neutralino
relic abundance.  Thus an upper bound on $\ohsq$ translates into an
upper bound on the parameters $m_0, \m12$.  The experimental and
cosmological constraints are nicely complementary, in part because the
cosmological bounds put upper limits on the parameters for which
particle searches provide lower limits.

The dark shaded area in Fig.~\ref{fig:bnds} delimits the
cosmologically preferred region with $0.1<\ohsq<0.3$.  The upper
limit, as described above, comes from an upper bound on the age of the
universe.  The lower limit is more of a preference than a
bound, stemming from the wish to have the neutralinos comprise
a significant fraction of the dark matter.  The two narrow vertical
channels arise from s-channel neutralino annihilation on the Higgs
and $Z^0$ poles (for some values of $\tb$ they have
merged into one large pole region).  Note that the top of the shaded area
intersects with the ``theory'' excluded area (where a stau is the LSP)
at $\m12\sim450$\gev.  Thus $\ohsq<0.3$ yields an upper bound on
$\m12$ and on $m_0$ (except in the close vicinity of the pole region,
which is largely excluded by the LEP chargino searches).  At
sufficiently low $\tb$, the Higgs bound moves to the right of the
shaded region, and the incompatibility of a Higgs lower bound of
$86\gev$ with the lower limit on the age of the universe excludes
values of $\tb$ less than 2, for $\mu<0$, or 1.65 for $\mu>0$
\cite{efgos}.

The dark solid line in Fig.~\ref{fig:vevs} is reproduced in
Fig.~\ref{fig:bnds}, where we have chosen $A_0=-\m12$ to minimise the
size of area containing CCB minima (see above and Ref.\cite{us0}).  We
note that there is only a restricted region in $\{\m12,m_0\}$ which is
both cosmologically and experimentally viable and which is free of CCB
minima, and moreover, such a region exists only for $\tb>2.3$ for
$\mu<0$ and $\tb>2.1$ for $\mu>0$.  At large $\tb\geqsim20$, \mbox{s-channel}
annihilation through the pseudoscalar Higgs and heavy Higgs can
contribute significantly to the neutralino annihilation cross-section,
and the above cosmological bounds are weakened\cite{bk1}.
Coannihilations with light sleptons also substantially reduces the
relic abundance of neutralinos near the line $m_{\tilde\tau}=\mchi$
and can provide a window at large $\m12$ \cite{efo}.

\section{Conclusions}

In this paper we have re-examined the unphysical charge and colour
breaking minima in the Constrained MSSM. In summary, we find that most
of the parameter space which is not already excluded by experiment or
by cosmological considerations has an unphysical vacuum which is lower
than the physical one. The region of parameter space which remains is roughly
\begin{eqnarray}
110 \leqsim &\m12& \leqsim 400\gev \nn\\
80 \leqsim  &m_0& \leqsim 170\gev ,
\end{eqnarray}
although there is a narrow region at $\m12\approx 150\,(110)\gev$ for
$\tb=10 \,(3)$ for which higher values of $m_0$ are still allowed by
the projected chargino searches.  Future tri-lepton searches at the
Tevatron should push the lower bound on $\m12$ to from $180-240\gev$
for the regions of Fig.~\ref{fig:bnds} \cite{bk2} and will help
close this loophole to larger $m_0$.

The regions which do have an unphysical minimum, as we stated in the
introduction, are not completely excluded but they must have a
constrained cosmology. Some general points were made in
Refs.\cite{riotto,us0}, so let us apply some of these considerations
to this specific model. Our main requirement for any acceptable
cosmological scenario is that it should explain why the physical
minimum is chosen instead of the global minimum. There are currently
three possible explanations;
\begin{itemize}
\item A high reheat temperature after inflation
\item Extra contributions to the effective potential during inflation
\item Breaking of $R$-parity 
\end{itemize}

\begin{figure}
\vspace*{-2.3in}
\begin{minipage}{6.0cm}
\hspace*{-1in}
\epsfig{file=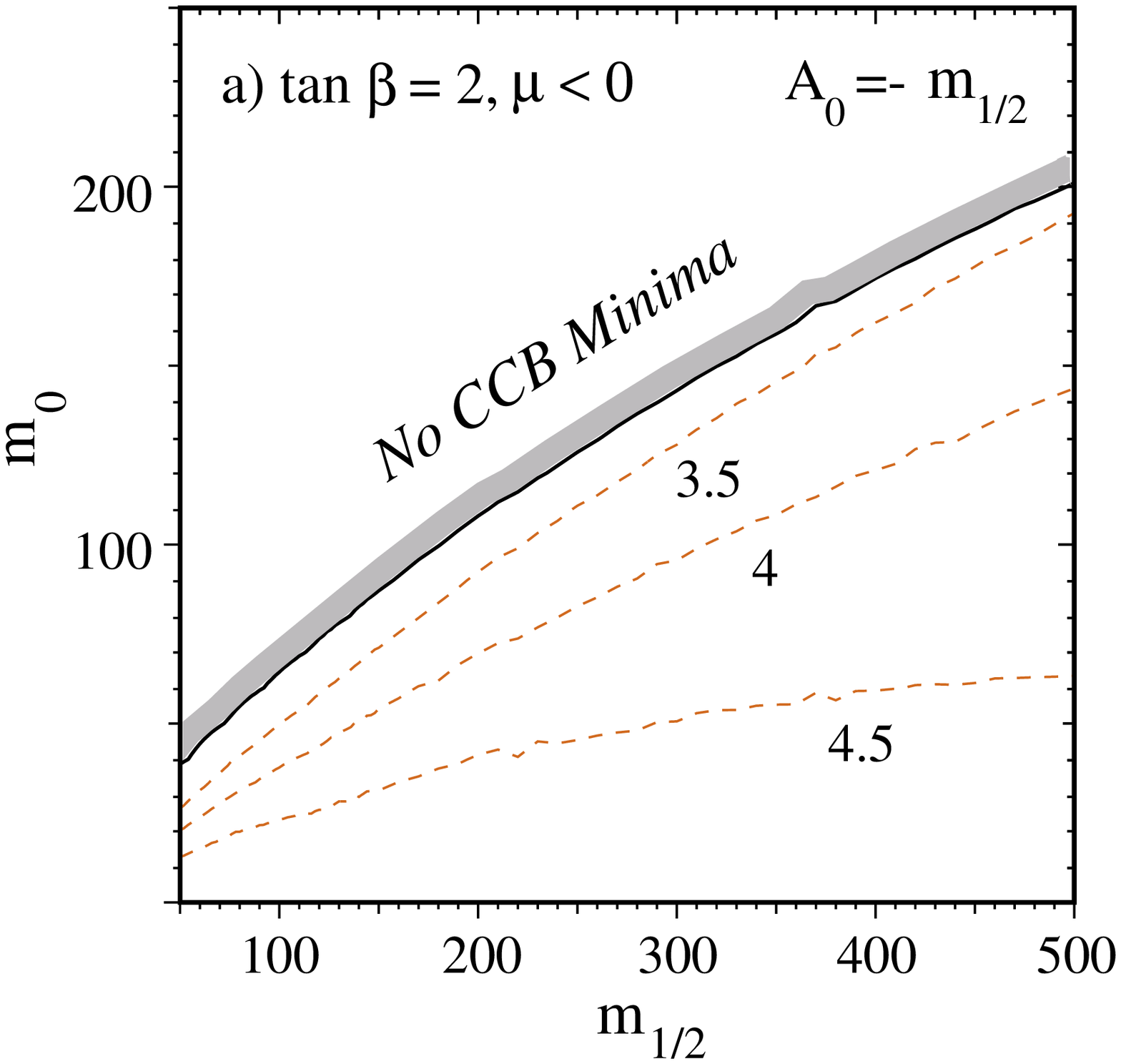,height=6in} 
\end{minipage}
\hspace*{0.3in}
\begin{minipage}{6.0cm}
\epsfig{file=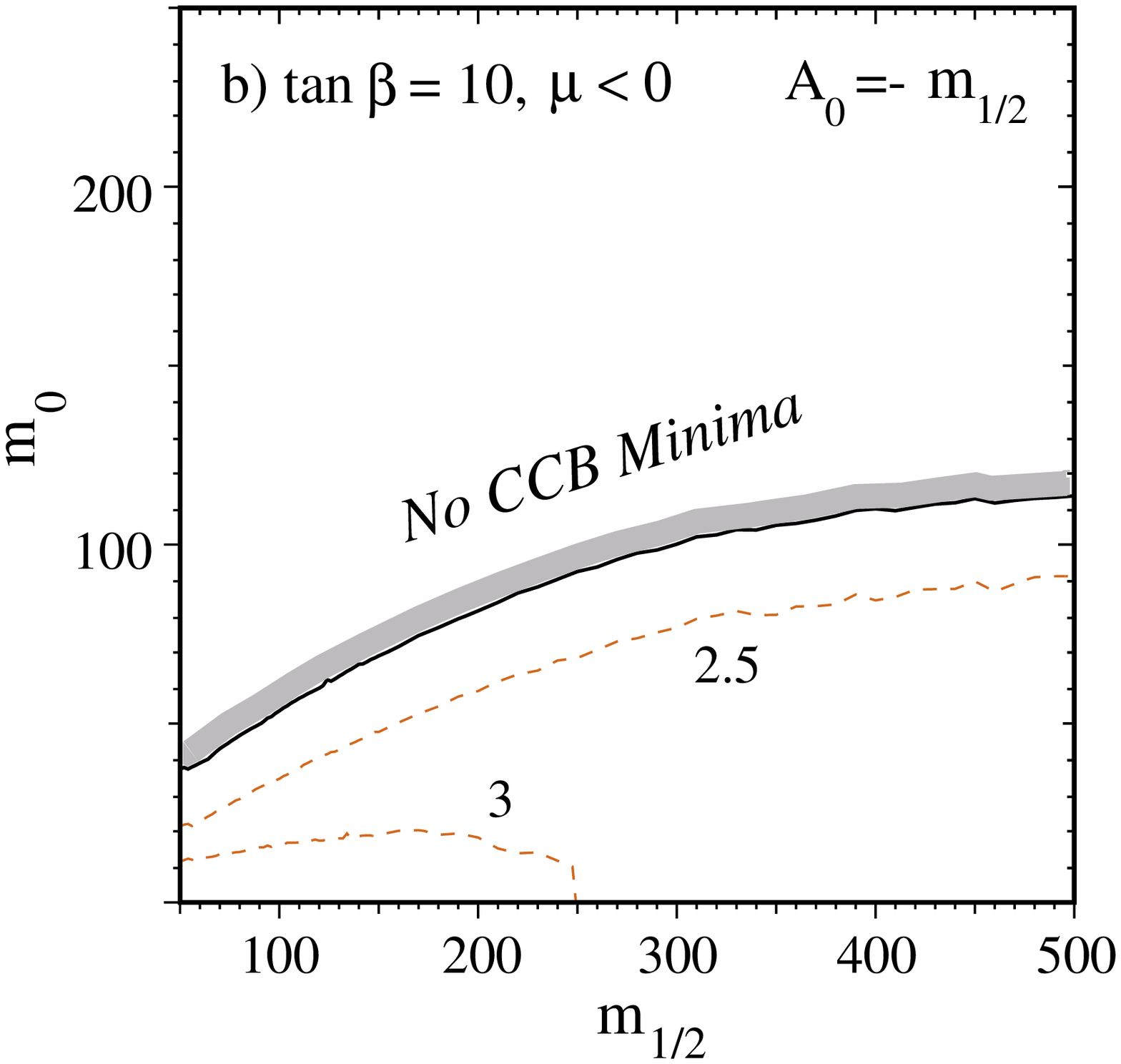,height=6in} 
\end{minipage}\hfill
\vspace{-2.0in}
\begin{minipage}{6.0cm}
\hspace*{-1in}
\epsfig{file=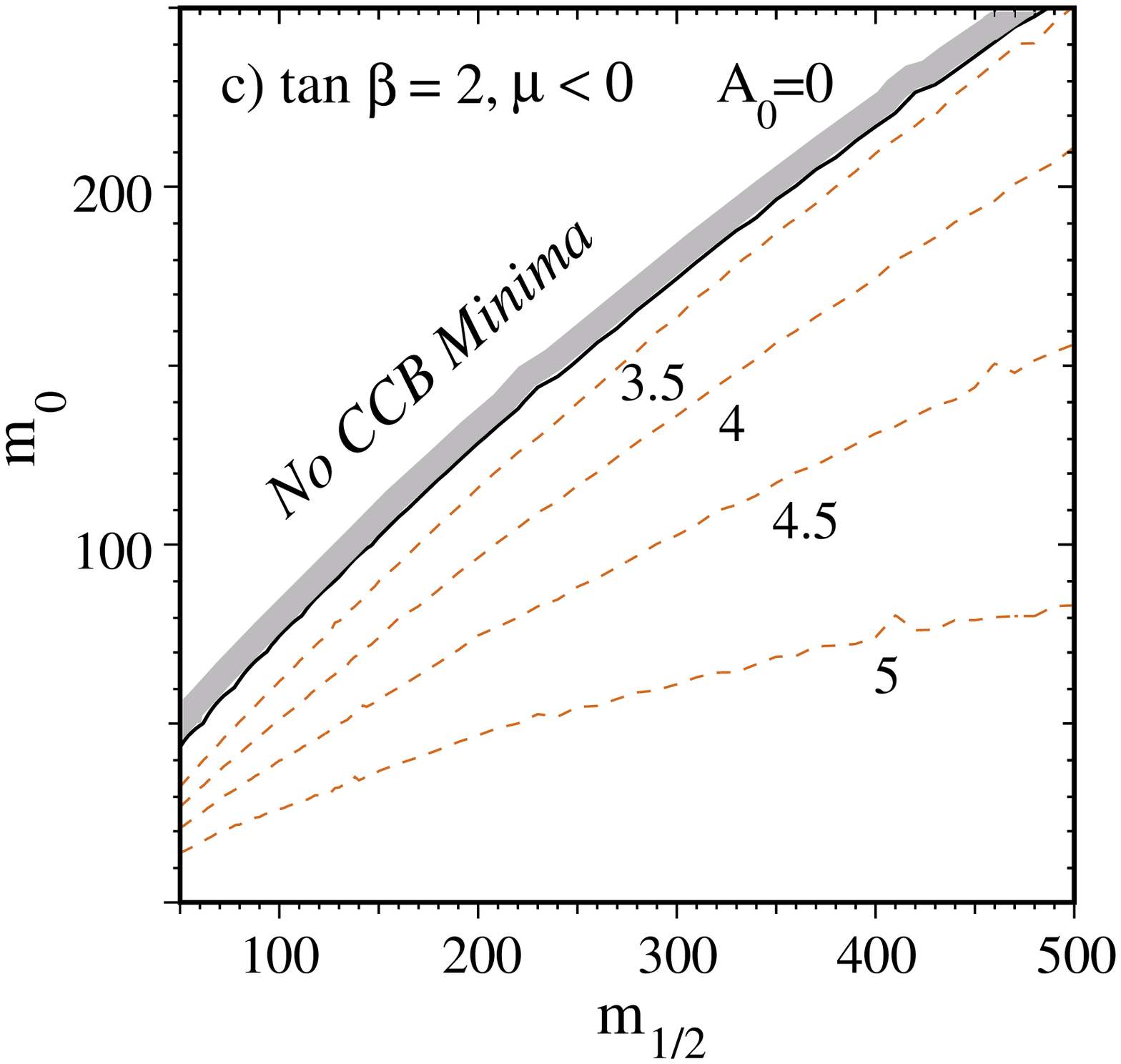,height=6in} 
\end{minipage}
\hspace*{0.3in}
\begin{minipage}{6.0cm}
\epsfig{file=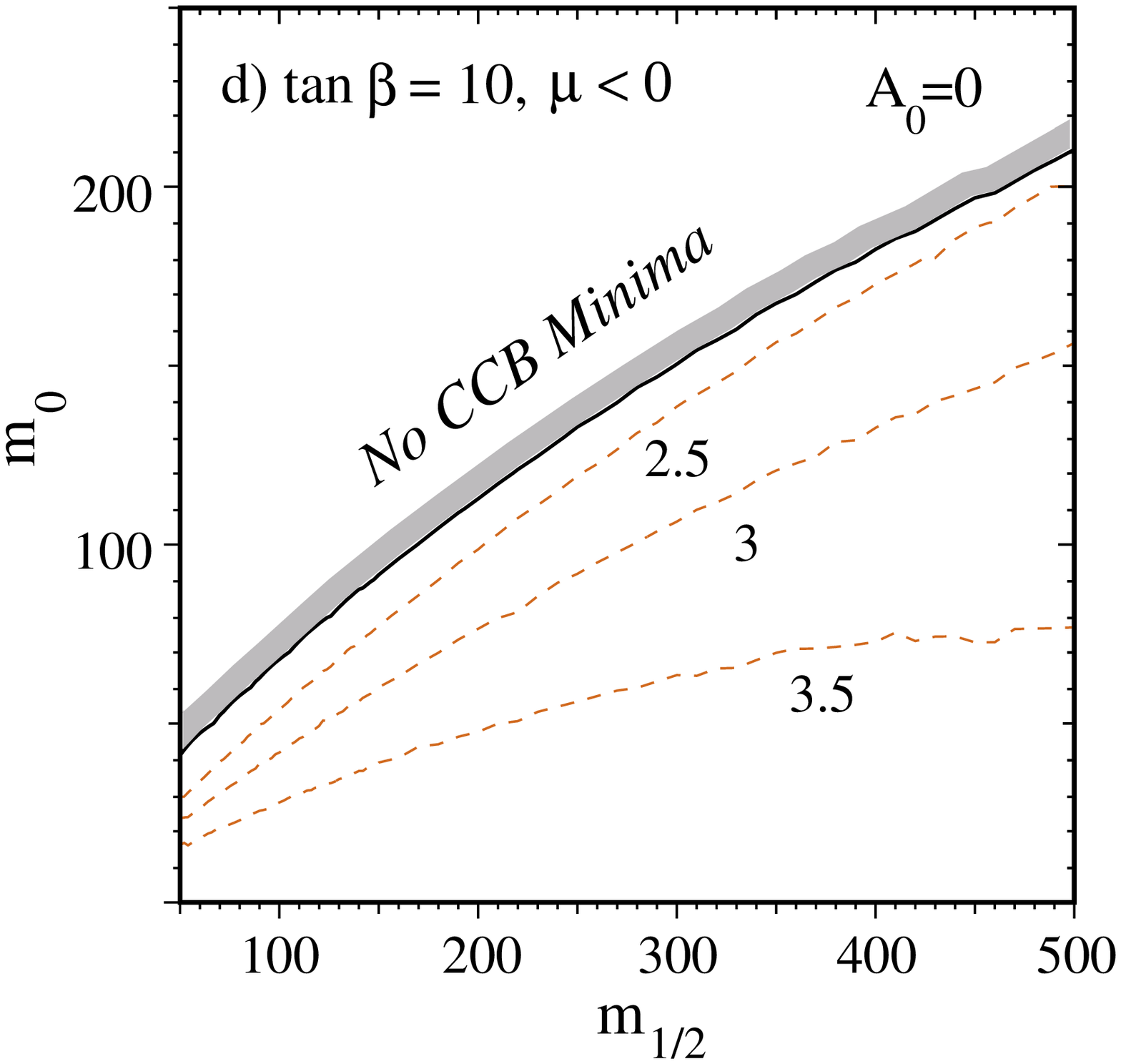,height=6in} 
\end{minipage}\hfill
\vspace{-0.7in}
\caption{\label{fig:vevs}The location of the CCB VEV, in the direction
$LLE, LH_2$, for $\tb=2,10$ and $A_0=0,-\m12$.  The dashed lines are
contours of constant $\log_{10}({\rm VEV/246}\gev)$.  In the shaded
strip, the CCB minimum is not global.  There are no CCB minima above
the shaded region.}
\end{figure}

In the case that the field is driven to the origin during reheating,
there is an additional constraint in all models in which supersymmetry 
breaking is transmitted to the physical sector gravitationally. In these 
models the gravitino mass is typically of the same order as the weak scale, 
and successful nucleosynthesis requires that 
\begin{equation}
\label{tr}
T_{reheat} \leqsim 10^9\gev .
\end{equation}
In order to lift the unphysical minimum we should have a reheat
temperature which is greater than roughly $10^{-1\pm 1}\times$ the
scale of the VEV (the `uncertainty' coming from the shape of the
effective potential and the contribution of light particles to
it). Hence a particular $T_{reheat}$ can be effective in the region
bounded by the corresponding contour in Fig.~(\ref{fig:vevs}).  From
Fig.~(\ref{fig:vevs}) we see that the condition (\ref{tr}) can
quite easily be satisfied away from the fixed point in the CMSSM. 

If the physical vacuum is chosen because of extra terms generated during 
inflation then there are constraints on the possible inflationary potentials. 
For example potentials with a Heisenberg symmetry are eliminated in this 
case because they are flat at tree level and the one loop corrections to the 
mass-squareds tend to be negative~\cite{BG,gmo}.

If $R$-parity is broken explicitly~\cite{quasi,us0} we must choose a
symmetry other than $R$-parity to prevent the proton from decaying,
and of course lose the neutralino as a dark matter candidate (but, on
the plus side, free up a large region of parameter space in
Fig.~(\ref{fig:bnds})).  Note that, as well as the lepton number
violating version of $R$-parity violation considered in
Ref.\cite{us0}, the CCB minima would also be lifted by a large 
Majorana mass for a right handed neutrino.  If
$R$-parity is broken during a stage of pre-heating~\cite{riotto2} then
it is in principle possible to have neutralino dark matter which is
stable. It is also possible that further lifting of the potential 
would occur in any case from extra operators required in the 
visible sector.  Both of these mechanisms
work because at large field values (\ie those corresponding to the VEV
of the unphysical minimum), the MSSM is no longer a good description
of the effective potential, and the potential is no longer $F$ flat 
along the relevant directions.

\section{Acknowledgements}
We would like to thank Sasha Davidson and Carlos Savoy for discussions, 
and the CERN theory division for hospitality.

\newpage 

\newpage 

\begin{thebibliography}{99}

\bibitem{ccb1}
J.-M. Fr\`ere, D.R.T. Jones and S. Raby, \npb{222}{11}{1983};
M. Claudson, L. Hall and I. Hinchcliffe, \npb{228}{501}{1983};
H.-P. Nilles, M. Srednicki and D. Wyler, 
\plb{120}{346}{1983};
J-P. Derendinger and C.A. Savoy, \npb{237}{307}{1984};
H. Komatsu, \plb{215}{323}{1988};
P. Langacker and N. Polonsky, \phrd{50}{2199}{1994};
A. Kusenko, P. Langacker and G. Segre, \prd{54}{5824}{1996};
J.A. Casas and S. Dimopoulos, \plb{387}{107}{1996};
J.A. Casas, \hepph{9707475};
J.A. Casas, A. Lleyda and C. Munoz, Nucl. Phys. {\bf B471} (1996) 3;
J.A. Casas, A. Lleyda and C. Munoz, \plb{389}{305}{1996};
I. Dasgupta, R. Rademacher and P. Suranyi, \hepph{9804229}

\bibitem{dilaton}
J.A. Casas, A. Lleyda and C. Munoz, \plb{380}{59}{1996}

\bibitem{baer}
H. Baer, M. Brhlik, D. Castano, \prd{54}{6944}{1996}

\bibitem{riotto}
T. Falk, K.A. Olive, L. Roszkowski and 
M. Srednicki, \plb{367}{183}{1996};
A. Riotto and E. Roulet, \plb{377}{60}{1996};
A. Strumia, \npb{482}{24}{1996};
T. Falk, K.A. Olive, L. Roszkowski, A. Singh and 
M. Srednicki, \plb{396}{50}{1997}

\bibitem{quasi}
S.A. Abel and B.C. Allanach, \plb{415}{1997}{371}; \hepph{9803476}

\bibitem{us0}
S.A. Abel and C.A. Savoy, \hepph{9803218}

\bibitem{us}
S.A. Abel and C.A. Savoy, \hepph{9809498}

\bibitem{recent}
J.A. Casas, A. Ibarra and C. Munoz, \hepph{9810266}

\bibitem{BG} P.~Binetruy and M.K.~Gaillard, 
\plb{195}{382}{1987}

\bibitem{gmo} M.-K. Gaillard, H. Murayama, and K. Olive,
\plb{355}{71}{1995}

\bibitem{carlos}
F. Buccella, J-P. Derendinger, C. A. Savoy and 
S. Ferrara, \plb{115}{375}{1982}; R. Gatto and G. Sartori,
\plb{157}{389}{1985}; M. A. Luty and W. Taylor, 
\prd{53}{3399}{1996}

\bibitem{tony}T. Ghergetta, C. Kolda, S. P. Martin, 
\npb{468}{37}{1996}

\bibitem{alephslps}ALEPH Collaboration, R. Barate et al., 
CERN preprint EP/98-077 (1998)

\bibitem{MSSMHiggs}H.E. Haber, R. Hempfling and A.H. Hoang,
Z. Phys.  {\bf C75} 539;
see also M. Carena, M. Quiros and C.E.M. Wagner, 
\npb{461}{407}{1996}

\bibitem{gold}H. Goldberg, Phys. Rev. Lett. {\bf 50} (1983) 1419

\bibitem{susydm1} J. Ellis, J.S. Hagelin, D.V. Nanopoulos, K.A. Olive
and M. Srednicki,  Nucl. Phys. {\bf B238} (1984) 453

\bibitem{efgos} J. Ellis, T. Falk, G. Ganis, K.A. Olive and
M. Schmitt, Phys. Rev. {\bf D58} (1998)  095002.

\bibitem{bk1} M. Drees and M. M. Nojiri, Phys. Rev. {\bf D47} (1993) 376; 
V. Barger and C. Kao, Phys. Rev. {\bf D57} (1998) 3131.

\bibitem{efo} J. Ellis, T. Falk, and K.A. Olive, \hepph{9810360}

\bibitem{bk2} V. Barger and C. Kao, MAD-98-1085, in preparation.

\bibitem{riotto2}
A. Riotto, E Roulet and I Vilja, 
\plb{390}{73}{1997}

\end{thebibliography}
\end{document}